\newtheorem{theorem}{Theorem}
\newcommand{\norm}[1]{\left\lVert#1\right\rVert}
\newcommand\abs[1]{\left|#1\right|}

\documentclass[journal]{IEEEtran}
\ifCLASSINFOpdf
   \usepackage[pdftex]{graphicx}
\else
   \usepackage[dvips]{graphicx}
\fi
\usepackage[cmex10]{amsmath}
\usepackage{amssymb}

\begin{document}
%
\title{Hard Threshold Least Mean Squares Algorithm}
%
%
%
\author{Lampros~Flokas
        and~Petros~Maragos
\thanks{The authors are with the National Tech. University of Athens, School of ECE, Greece, Email:lamflokas@gmail.com, maragos@cs.ntua.gr}
}
\maketitle

\begin{abstract}
This work presents a new variation of the commonly used Least Mean Squares Algorithm (LMS) for the identification of sparse signals with an a-priori known sparsity using a hard threshold operator in every iteration. It examines some useful properties of the algorithm and compares it with the traditional LMS and other sparsity aware variations of the same algorithm. It goes on to examine the application of the algorithm in the problem of spectrum estimation for cognitive radio devices.   
\end{abstract}

\begin{IEEEkeywords}
signal processing, sparse representations, LMS, cognitive radio.
\end{IEEEkeywords}

%
\IEEEpeerreviewmaketitle

\section{Introduction}
%
%
%
%
\IEEEPARstart{L}{east} Mean Squares Algorithm (LMS), introduced by Widrow and Hoff \cite{WidrowAd}, is an algorithm used in many signal processing tasks like adaptive system identification. Of course, the LMS algorithm is not optimized to take advantage of special features of the estimated vector. Taking under consideration prior knowledge of the estimated vector can allow us to achieve  faster convergence, a smaller steady state error or reduced time complexity. Although such prior knowledge tends to be application specific and techniques used may vary widely in a case by case basis, one commonly used property of the estimated vector is sparsity. Sparsity and its applications has been thoroughly studied in \cite{Cand05} and \cite{Dono02} and therefore a variety of algorithms has been introduced. There has been a lot of scientific work like \cite{Them14} on the case of adaptive algorithms but most of the algorithms developed do not have the simplicity and the time complexity of the LMS algorithm.     
\par In order to overcome this drawback  \cite{Chen09}  introduces variations of the LMS algorithm in order to induce sparsity on the estimated vector. Here we will propose a new variation, named Hard Threshold LMS of the LMS algorithm that alternates standard LMS update with shrinking using a hard threshold operator. This algorithm is the adaptive version of the iterative hard thresholding studied in \cite{Blum09} and \cite{Blum10}.
\par The structure of the paper is as follows. In Section 2 we discuss the properties of Hard Threshold LMS and some variations of it. In Section 3 numerical simulations comparing Hard Threshold LMS with other sparsity aware algorithms are provided. In Section 4 we discuss the application of the algorithm proposed for the problem of spectrum estimation for cognitive radio devices. Finally, Section 5 contains concluding remarks and discusses possible directions for future research.
\vspace*{-8mm}
\section{Algorithms}
\subsection{The LMS algorithm}
The hard threshold LMS algorithm consists of alternating one step of traditional LMS update with a shrinkage step using a hard threshold operator. To better understand the procedure we review the first step. Let $y(n)$ be a sequence of observations of the output of a system following the model
\begin{equation} \label{eq:lms}
y(n)=\mathbf{w}^T\mathbf{x}(n)+v(n)
\end{equation}
where $\mathbf{w}=[w_0,w_1,\dots,w_{N-1}]^T \in \mathbb{R}^N$  is the parameter vector to be estimated, $\mathbf{x}(n)=[x(n),x(n-1),\dots,x(n-N+1)]^T \in \mathbb{R}^N$ consists of the last $N$ values of the system input and $v(n)$ is the additive observation noise. Let also $\mathbf{w}(n)$ be the estimation we have up to time $n$ for the unknown vector $\mathbf{w}$ and $e(n)$ be the approximation error. Then
\begin{equation}
e(n)=y(n)-\mathbf{w}^T(n)\mathbf{x}(n)
\end{equation}
\par The LMS update rule is recursive and produces a new estimation given the previous one, following the rule
\begin{equation}
\mathbf{w}(n+1)=\mathbf{w}(n)+\mu e(n)\mathbf{x}(n)
\end{equation}
where $\mu$ is a an appropriately chosen constant. If $\mathbf{R}_x$ is the covariance matrix of $\mathbf{x}(n)$ and $\lambda_{max}$ is its maximum eigenvalue then LMS will converge in the mean sense if:
\begin{equation} \label{eq:bounds}
0<\mu<1/\lambda_{max}
\end{equation}
\vspace*{-9mm}
\subsection{Hard Threshold LMS} 
Hard threshold LMS goes beyond that update rule using the shrinkage step. In order to do so however, an upper bound on the sparsity of the vector under estimation must be known in advance. Let $\mathrm{support}(\mathbf{x})=\{ i \in \{0,1,..,N-1\}: x_i \neq 0\}$ and $\norm{\mathbf{x}}_0=|\mathrm{support}(\mathbf{x})|$, where $|S|$ denotes the cardinality of set $S$. Also assume that we know that $\norm{\mathbf{w}}_0\leq s$ where $s$ is a positive integer less than $N$. Then the update rule of the Hard Threshold LMS becomes
\begin{equation} \label{eq:hard_lms}
\mathbf{w}(n+1)=H_s(\mathbf{w}(n)+\mu e(n)\mathbf{x}(n))
\end{equation}
where $H_s$ is the operator that outputs a vector having zeros in all coefficients except for the ones with the $s$ largest absolute values that remain the same as in the input vector. For example if $\mathbf{x}_0=[2,-2,1,0]^T$ then $H_2(\mathbf{x}_0)=[2,-2,0,0]^T$. In case of ties we can take a conservative approach and allow all tying coefficients to be nonzero in the resulting vector so that $H_1(\mathbf{x}_0)=[2,-2,0,0]^T$. Thus $|\mathrm{support}(H_s(\mathbf{x}))| \geq s$ and therefore it is not guaranteed that the output will always be $s$-sparse. The operator can give non $s$-sparse results when there are multiple coefficients in the vector that their absolute value is equal to the $s$ largest absolute value in the vector. However, in most cases such ties will be nonexistent and the result will be an $s$-sparse vector.
\par It is easy to see the similarity of our proposed algorithm with the iterative hard thresholding algorithm studied in \cite{Blum09} and \cite{Blum10}. There, since the algorithm is developed in a batch setting where all the data are known in advance, the relation between the observations $\mathbf{y}$ and the estimated vector $\mathbf{w}$ is $\mathbf{y}=\mathbf{A}\mathbf{w}$ where $\mathbf{A}$ is $M\times N$ matrix with $M<N$; thus the problem is underdefined. The update of the iterative hard thresholding under similar assumptions for the sparsity of $\mathbf{w}$ is
\begin{equation}
\mathbf{w}(n+1)=H_s(\mathbf{w}(n)+\mu \mathbf{A}^T\mathbf{e}(n))
\end{equation}
where $\mathbf{e}(n)=\mathbf{y}-\mathbf{A}\mathbf{w}(n)$.
\par As a result it is clear that the proposed algorithm is closely related to the special case of iterative hard thresholding having $M=1$. It is also clear that we cannot use the rigorous proofs found in \cite{Blum09} and \cite{Blum10} to show that the proposed algorithm also converges since for $M=1$ it is impossible to fulfil the strict properties needed. However, it is still possible to prove some interesting properties of the hard threshold operator. The main contribution of the operator is to let us focus our attention on the support of the estimated vector. If the algorithm does not provide a correct estimation of the support of the estimated vector then this could have a negative effect on the convergence of the algorithm. So one of the key properties that need to be studied is under which conditions is the estimation of the support using the hard threshold operator correct.
\begin{theorem} \label{th:strict}
Let $\mathbf{w}=[w_0,w_1,\dots,w_{N-1}]^T \in \mathbb{R}^N$ with $\norm{\mathbf{w}}_0=s$ and $\hat{\mathbf{w}}$ be an approximation. Let $q=\min_{w_i \neq 0} \abs{w_i}$. Then if $\norm{\mathbf{w}-\hat{\mathbf{w}}}_2^2 < \frac{q^2}{2}$ the following will be true 
\begin{equation} \label{eq:toprove}
\mathrm{support}(H_s(\hat{\mathbf{w}}))  = \mathrm{support}(\mathbf{w})
\end{equation}
\end{theorem}
\begin{IEEEproof}
The proof will be completed in three distinct cases.
\par (i) First, we assume that $\norm{H_s(\hat{\mathbf{w}})}_0<s$ which can be true only if $\norm{\hat{\mathbf{w}}}_0<s$. We can easily see that, since  $\norm{\mathbf{w}}_0=s$, there is at least one coefficient index $i$ such that $\hat{w}_i=0$ and $w_i\neq 0$, which from the hypothesis also means that $\abs{w_i}\geq q$. As a result we have that $$\norm{\mathbf{w}-\hat{\mathbf{w}}}_2^2 \geq \abs{w_i-\hat{w}_i}^2= \abs{w_i}^2 \geq q^2$$ which contradicts the hypothesis; so this case is impossible.
\par (ii) Now we have that $\norm{H_s(\hat{\mathbf{w}})}_0 = s$. Let us assume that relation (\ref{eq:toprove}) does not hold. Then since the two sets have the same number of nonzero elements, it is clear that there is a coefficient index $\ell\in \mathrm{support}(\mathbf{w})$ but $\ell \notin \mathrm{support}(H_s(\hat{\mathbf{w}}))$ and a coefficient index $k$ so that $k \in \mathrm{support}(H_s(\hat{\mathbf{w}}))$ but $k \notin \mathrm{support}(\mathbf{w})$. We directly know that $w_k=0$ and that $\abs{w_\ell}\geq q$. We can also deduce that $\abs{\hat{w}_k} > \abs{\hat{w}_\ell}$ since $k$ belongs in $\mathrm{support}(H_s(\hat{\mathbf{w}}))$ but $\ell$ does not. Then, for the error norm we have $$\norm{\mathbf{w}-\hat{\mathbf{w}}}_2^2 \geq \abs{w_k-\hat{w}_k}^2+\abs{w_\ell-\hat{w}_\ell}^2$$ Since $\abs{w_k-\hat{w}_k}^2 = \abs{\hat{w}_k}^2 > {\hat{w}_\ell}^2 $, it follows that $$\norm{\mathbf{w}-\hat{\mathbf{w}}}_2^2 > 2{\hat{w}_\ell}^2-2w_\ell \hat{w}_\ell + {w_\ell}^2 $$ Therefore we can also write that $$\norm{\mathbf{w}-\hat{\mathbf{w}}}_2^2 >\min_{\hat{w}_\ell \in \mathbb{R}}{2{\hat{w}_\ell}^2-2w_\ell \hat{w}_\ell + {w_\ell}^2} $$ The minimum value of the RHS is attained for $\hat{w}_\ell=\frac{w_\ell}{2}$ and equals $\frac{{w_\ell}^2}{2}$; hence $$\norm{\mathbf{w}-\hat{\mathbf{w}}}_2^2 >\frac{{w_\ell}^2}{2} \geq \frac{q^2}{2} $$ This once again contradicts the hypothesis and so relation (\ref{eq:toprove}) is true in this case.
\par (iii) Finally, we assume that $\norm{H_s(\hat{\mathbf{w}})}_0 > s$. This can happen only if there are ties for the s largest absolute value in $\hat{\mathbf{w}}$. Let us denote as $B$ the set of tying coefficients, $A= \mathrm{support}(H_s(\hat{\mathbf{w}})) \setminus B$ and finally $C=(\mathrm{support}(H_s(\hat{\mathbf{w}}))^c$. It is evident that $\abs{A} \leq s-1$. We shall prove that this case is impossible. There are two subcases:
\par (a) $B \cap \mathrm{support}(\mathbf{w})= \emptyset$. Since $\abs{A} \leq s-1$ and $\norm{w}_0=s$, $\mathrm{support}(\mathbf{w})$ must have an element in common with $C$. Let us call that element $\ell$. Let us also take an element $k$ from $B$.  Then just like in the second case $\abs{\hat{w}_k} > \abs{\hat{w}_\ell}$ since $k$ belongs in $\mathrm{support}(H_s(\hat{\mathbf{w}}))$ but $\ell$ does not. Following the rest of the steps in case (ii) we reach a contradiction.
\par (b) $B \cap \mathrm{support}(\mathbf{w})\neq \emptyset$. Let $\ell$ a common element of the two sets. Since $\norm{H_s(\hat{\mathbf{w}})}_0 >\norm{\mathbf{w}}_0$ there is an element $k$ so that $k \in \mathrm{support}(H_s(\hat{\mathbf{w}}))$ but $k \notin \mathrm{support}(\mathbf{w})$. Since $\ell$ is one of the indexes tying for the last spot, we have $\abs{\hat{w}_k} \geq \abs{\hat{w}_\ell}$. Following the steps of case (ii) yields $\norm{\mathbf{w}-\hat{\mathbf{w}}}_2^2 \geq \frac{{w_\ell}^2}{2} \geq \frac{q^2}{2} $ and therefore we get a contradiction.
\end{IEEEproof}
\par In order to understand the significance of the theorem we need to see some equivalent bounds having to do with the signal to error ratio that is needed so that the result in relation (\ref{eq:toprove}) still holds. The true vector $\mathbf{w}$ has $s$ nonzero values  each with an absolute value of at least $q$. Thus $\norm{\mathbf{w}}_2^2\geq sq^2$ and hence we need
\begin{equation} \label{eq:ser}
\mathrm{SER}=\frac{\norm{\mathbf{w}}_2^2}{\norm{\mathbf{w}-\hat{\mathbf{w}}}_2^2}> \frac{sq^2}{\frac{q^2}{2}}=2s
\end{equation}
\par Inequality (\ref{eq:ser}) is a necessary condition so that the required conditions of the theorem are true. Even if it is not sufficient it gives us the intuition that for small values of $s$ it will be easier to come up with an estimate $\hat{\mathbf{w}}$ for which relation (\ref{eq:toprove}) is true. On the other hand the conditions of Theorem~\ref{th:strict} are just sufficient for the relation (\ref{eq:toprove}) so in practice relation (\ref{eq:toprove}) could be true even with much lower signal to error ratios.
\par To further relax the conditions of our theorem we could allow the estimate to be less sparse. In order to do this we could use $H_d$ instead of $H_s$ with $N>d>s>0$ where $N$ is the size of the estimated vector. What happens here is a trade off. On the one hand, the result now is less attractive since we have more nonzero coefficients than what is actually needed and that may lead to excessive estimation error that could possibly be avoided. On the other hand, the estimation error of the input to the threshold operator can be greater without risking of loosing an element of $\mathrm{support}(\mathbf{w})$ after the application of the operator. The next theorem quantifies the gain in allowable estimation error.
\begin{theorem} \label{th:relaxed}
Let $\mathbf{w}$ be a vector in $\mathbb{R}^N$ with $\norm{\mathbf{w}}_0=s$ and $\hat{\mathbf{w}}$ be an approximation. Let $q=\min_{w_i \neq 0} \abs{w_i}$ and $d=s+\tau$ with $d<N$ and $\tau>0$ where $s$, $\tau$, $d$ are integers. Then if $\norm{\mathbf{w}-\hat{\mathbf{w}}}_2^2\leq q^2(1-\frac{1}{\tau+2})$ and $\norm{\hat{\mathbf{w}}}_0 \geq d $, the following will be true
\begin{equation} \label{eq:toprove_2}
\mathrm{support}(H_d(\hat{\mathbf{w}}))  \supseteq \mathrm{support}(\mathbf{w})
\end{equation} 
\end{theorem}
\par The analogous inequality of relation (\ref{eq:ser}) for this theorem, whose proof can be found in appendix \ref{ap:relaxed}, can be found as
\begin{equation}
\mathrm{SER} \geq \frac{s}{(1-\frac{1}{\tau+2})}
\end{equation}
which is less strict as we have expected.
\par Given the last theorem one can overcome the need to have an initialization $\mathbf{w}(0)$ such that $\norm{\mathbf{w}(0)-\hat{\mathbf{w}}}_2$ is small in order to potentially avoid losing coefficients of $\mathrm{support}(\hat{\mathbf{w}})$.
\subsection{Selective Zero-Attracting LMS }
\par One more way to overcome the need to have a small initial error, $\norm{\mathbf{w}(0)-\hat{\mathbf{w}}}_2$, and still enforce sparsity is to abandon the hard threshold operator and further relax the conditions of convergence. One idea is instead of assigning a zero to coefficients that are deemed to be unnecessary, one could penalize them by reducing their absolute value by a constant $\rho$. This is the same concept of the $\ell_1$ penalization presented in \cite{Chen09} but applied only to the possibly superfluous coefficients given the a priori estimation of sparsity. Then the update rule of every coefficient will be
\begin{align}
\mathbf{u}(n)& = \mathbf{w}(n)+\mu e(n)\mathbf{x}(n)\\
w_{i}(n+1)&=
\begin{cases}
u_{i}(n) ,\quad i \in \mathrm{support}(H_s(\mathbf{w}(n))) \\
u_{i}(n) -\rho \mathrm{sgn}(w_{i}(n)), \text{ otherwise}
\end{cases}
\end{align}
where $\mathbf{u}(n)$ with elements $u_i(n)$ is the vector corresponding to the simple LMS update and $\mathrm{sgn}(x)=\frac{x}{\abs{x}}$ if $x \neq 0$ and zero else. For simplification we can define a penalty operator $P_s$:
\begin{equation}
{P_s(\mathbf{x})}_{i}=
\begin{cases}
0, \quad i \in \mathrm{support}(H_s(x))\\
\mathrm{sgn}(x_i),  \text{ otherwise} 
\end{cases}
\end{equation}
so that the update rule can be written
\begin{equation} \label{eq:relaxed}
\mathbf{w}(n+1)=\mathbf{u}(n)-\rho P_s(\mathbf{w}(n))
\end{equation}
\par For this algorithm we can prove the following theorem
\begin{theorem} \label{th:sza-lms}
Let us have a zero mean observation noise $v(n)$ independent of $\mathbf{x}(n)$ and given that $\mathbf{x}(n)$ and $\mathbf{w}(n)$ are independent then the algorithm described by (\ref{eq:relaxed}) converges in the mean sense provided that the condition of (\ref{eq:bounds}) holds. The limiting vector satisfies the equation
\begin{equation} \label{eq:bias}
\mathbb{E}[\mathbf{w}(\infty)]=\mathbf{w}-\frac{\rho}{\mu}\mathbf{R}_x^{-1}\mathbb{E}[P_s(\mathbf{w}(\infty))]
\end{equation}
\end{theorem}
\par Even though the proof, found in the appexdix \ref{ap:sza-lms}, is similar to the proof of convergence of the Zero-Attracting LMS Algorithm (ZA-LMS) presented in \cite{Chen09}, the algorithm presented here is closer to the logic of the Reweighted version of ZA-LMS (RZA-LMS) presented also in \cite{Chen09}. In RZA-LMS the penalty introduced in all coefficients is inversely proportionate to the coefficients magnitude. In the algorithm presented here coefficients that are large relative to others in the sense that they belong in $\mathrm{support}(H_s(\mathbf{w}(n)))$ are not penalized and all the other ones are penalized by a constant factor. The result is that according to the equation (\ref{eq:bias}) the bias of the algorithm is zero for the coefficients that are believed to be in the $\mathrm{support}(\mathbf{w})$. By using the previous theorems it is easy to see that we can choose a small enough $\rho$ to reduce the bias of estimation and guarantee a correct estimation of the support in the mean sense and thus zero bias in those coefficients. If the exact support is found then the bias is zero altogether.
\section{Experimentation}
\begin{figure}[!t]
\centering
\includegraphics[width=2.5in]{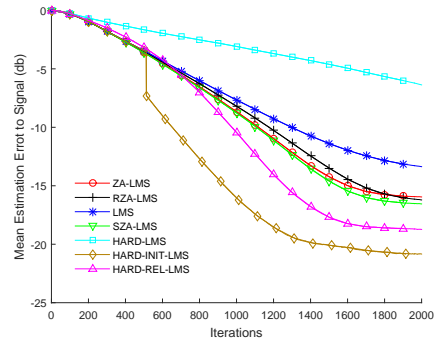}
\caption{Estimation of a 256 tap filter with 28 non zero taps. This figure is better to view in color.}
\label{fig:stationay}
\end{figure} 
In this section we will compare the performance of the various algorithms discussed previously. The first experiment uses the following setting: A signal of length 2000 consisting of samples drawn from the normal distribution is used as input for a filter with 256 taps, of which 28 are randomly selected to have the value 1 and the rest are set to zero. The output is then affected by additive white gaussian noise such that the resulting SNR is 30 db. For this estimation task we used algorithms presented in other works such as the standard LMS algorithm and the ZA-LMS, RZA-LMS algorithms as presented in \cite{Chen09}. Additionally we used the algorithms which were introduced in this paper: the Selective ZA-LMS (SZA-LMS) as presented in (\ref{eq:update}), the Hard Threshold LMS (HARD-LMS) as discussed in (\ref{eq:hard_lms}), a variation of HARD-LMS called here HARD-INIT-LMS where the first 512 updates do not use the hard threshold operator for better initialization and the relaxed version of HARD-LMS called here HARD-REL-LMS as discussed in Theorem \ref{th:relaxed}. The parameters used here are the following: $\mu=0.005$, $\rho=5 \times 10^{-5}$ , the $\epsilon$ parameter of RZA-LMS is set to 10, $s=28$ and $d=56$ for HARD-REL-LMS. The results shown in Figure \ref{fig:stationay} come from the mean of 200 executions of the experiment and depict the error to signal ratio of the estimation $\hat{\mathbf{w}}(n)$ in each iteration.
\par As shown in Fig.~\ref{fig:stationay}, the HARD-LMS algorithm fails the estimation task (or takes too many iterations to achieve a respectable error to signal ratio) as it is unable to find the correct support whereas its variations are able to track it just fine giving the best results among all algorithms. The algorithms that follow the Zero Attracting scheme perform better than the standard LMS and the SZA-LMS that we propose is the best performing among those.

\section{Cognitive radio application}
One of the advantages of using Compressive Sampling techniques is that one can use the a priori knowledge of sparsity to reduce the number of samples needed to estimate the unknown sparse vector. This property can be very useful when the number of samples needed would be prohibitively large. For example, in most applications that deal with wide band signals using the Nyquist Frequency to sample the input can be very costly. One such application is spectrum estimation for cognitive radio devices. Wireless communication spectrum is a limited resource so it is impossible to statically split the spectrum among all the possible applications. To overcome this limitation  cognitive radio devices try to dynamically manage the spectrum by detecting which part of the spectrum is unused by its primary users and temporarily use it for their own needs. In order to be effective these devices would need to check a wide band of frequencies to increase the possibility of finding unused spectrum. The high sampling frequency needed would increase the cost of such devices.
\par If it were possible to write the signal as a linear transformation of a sparse vector then we could leverage the techniques of Compressive Sampling to reduce the samples needed and increase the accuracy of our estimation. Generally this is dependent on the nature of the signal. However, in our case we can use the fact that the spectrum of the signal should be sparse as many of the frequencies will be left unused by its users. Let us define as $\mathbf{U}$ the undersampling matrix, whose rows are a subset of the rows of the identity matrix. Additionally, let us define $\mathbf{\Phi}$ as the matrix whose application on a row vector results in the Inverse Discrete Fourier Transform of the vector. Moreover $\mathbf{w}$ corresponds to the DFT of the signal which is sparse. Then the samples received can be written as 
\begin{equation}
\mathbf{y}=(\mathbf{U}\mathbf{\Phi})\mathbf{w}
\end{equation}
\par This equation formulates the batch version of problem which can be solved with the Compressive Sampling algorithms. This approach has also been studied in \cite{Tian07} and \cite{Tian08}. 
\par In order to use the adaptive algorithms in this case we need to make some adjustments. First of all, we are no longer simulating a FIR system. So the $\mathbf{x}(n)$ as seen in (\ref{eq:lms}) comes from the transpose of the rows of $\mathbf{U}\mathbf{\Phi}$. Additionally, the equations must be updated for the complex case:
\begin{equation*}
\begin{array}{c}
y(n)=\mathbf{w}^H\mathbf{x}(n)+v(n), \quad e(n)=y(n)-\mathbf{w}^H(n)\mathbf{x}(n) \\
\mathbf{w}(n+1)=\mathbf{w}(n)+\mu e^*(n)\mathbf{x}(n)
\end{array}
\end{equation*}
\par The hard threshold operator can be extended in the complex case by comparing the magnitudes of the complex values instead of the absolute values.
\par
\begin{figure}[!t]
\centering
\includegraphics[width=2.5in]{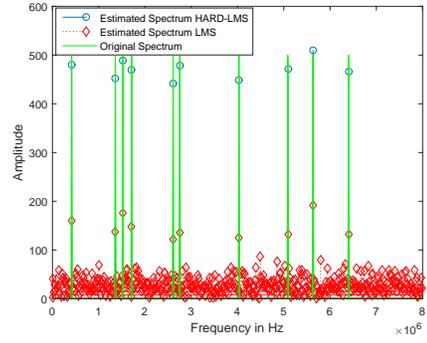}
\caption{Estimation of the spectrum of the undersampled superposition of 10 sine waves. This figure is better to view in color.}
\label{fig:radio}
\end{figure}

 In order to examine the performance of the Hard Threshold LMS algorithm we will do the following experiment: Firstly we take as our signal a superposition of 10 sine waves with 10 random frequencies matching the DFT frequency bins corrupted by additive white gaussian noise so that the resulting SNR is equal to 20 db. The length of our signal is 1000 samples. Then we take randomly 300 samples from it. It is important that the samples are taken randomly and not uniformly in order to avoid aliasing which leads to unrecoverable loss of information. We choose the value of $\mu$ to be the inverse of the squared norm of $\mathbf{x}(n)$ which is constant as $\mathbf{x}(n)$ are formed by DFT rows. The last problem we need to cover is that of the convergence of the algorithms. With only 300 samples our algorithms may not converge fast enough to our solution. The LMS algorithm and its variations are guaranteed to converge only when the number of iterations goes to infinity. To solve this problem we choose to retrain the estimation of both of this algorithms with the same data 10 times. Moreover, in order to have a good initial estimate our Hard Threshold LMS algorithm does not use thresholding during the first session. Moreover, $s$ is set equal to 20 as each sine will be represented as two non zero values in the frequency domain. As shown in Fig.~\ref{fig:radio} the estimation of the Hard Threshold LMS algorithm manages to track all the frequencies along with their respective amplitudes. The indexes of the $s$ values with the largest amplitude of the standard LMS algorithm estimate also track the significant frequencies. However, the corresponding amplitudes are much smaller than in the original spectrum and not very discernible from amplitudes computed for the rest of the frequencies where the original spectrum is zero. 
\vspace*{-2mm}  
\section{Conclusion}
We firstly examined a variation of the standard LMS algorithm where the hard threshold operator is used in between updates to enforce sparsity on the estimated vector. Additionally, we discussed the effectiveness of allowing our estimation to be less sparse in order to overcome the need of a good initial estimation. Moreover, we used the idea of the support estimation with the hard threshold operator in order to improve the performance of the ZA-LMS algorithm presented in \cite{Chen09}. Further, we presented the results of experiments that compare the various algorithms proposed here with the already existing ones. Finally, we discussed the problem of spectrum estimation for cognitive radio devices and how the underlying spectrum sparsity may be useful to achieve improved estimations even when the data is undersampled and noisy. 
\par Of course this is one of the many possible applications of the proposed algorithms. Obviously an a priori estimation of the sparsity of the estimated vector may not be available in all use cases, even though we showed that this estimate must not be exact in order to actually take benefit. However, there are other use cases where the algorithms proposed here could make a difference. The standard LMS algorithm has been used in many adaptive machine learning tasks like neural network training and others as discussed in \cite{Theod15} so taking advantage of sparsity could be advantageous. For example, in the case of training a perceptron with an abundance of available features one could begin training with all the features but then proceed to use one of the proposed algorithms to impose feature selection through sparsity. By increasing the imposed sparsity one can then train several classifiers and then compare them using criteria like the Bayesian information criterion.


%



\ifCLASSOPTIONcaptionsoff
  \newpage
\fi



%
\bibliographystyle{IEEEtran}
\bibliography{IEEEabrv,publication}

\appendices
\section{Proof of the Theorem \ref{th:relaxed}} \label{ap:relaxed}
\begin{IEEEproof}
Let us assume that relation (\ref{eq:toprove_2}) does not hold. Just like in the previous proof it is clear that there is a coefficient index so that $\ell\in \mathrm{support}(\mathbf{w})$ but $\ell \notin \mathrm{support}(H_d(\hat{\mathbf{w}}))$. This time however the set $\mathrm{support}(H_d(\hat{\mathbf{w}}))$ has at least $d=s+\tau$ elements but $\mathrm{support}(\mathbf{w})$ has at most $s-1$ elements that could exist in  $\mathrm{support}(H_d(\hat{\mathbf{w}}))$. As a result we are sure that there are at least $\tau+1$ indexes $k_i$ so that $k_i \in \mathrm{support}(H_s(\hat{\mathbf{w}}))$ but $k_i \notin \mathrm{support}(\mathbf{w})$. Once again we know that $w_{k_i}=0$ and that $\abs{w_\ell}\geq q$ and we can deduce that $\abs{\hat{w}_{k_i}} > \abs{\hat{w}_\ell}$ since $k_i$ exists in $\mathrm{support}(H_d(\hat{\mathbf{w}}))$ but $\ell$ does not.
\par Like in the previous proof we can deduce about the error norm that $$\norm{\mathbf{w}-\hat{\mathbf{w}}}_2^2 \geq \sum_{i=1}^{\tau+1}{\abs{w_{k_i}-\hat{w}_{k_i}}^2}+\abs{w_\ell-\hat{w}_\ell}^2$$
We bound the first term just like in the previous proof so that it becomes $$\sum_{i=1}^{\tau+1}{\abs{w_{k_i}-\hat{w}_{k_i}}^2} = \sum_{i=1}^{\tau+1}{\hat{w}_{k_i}^2} \geq (\tau+1) {\hat{w}_\ell}^2$$ Thus, we end up $$\norm{\mathbf{w}-\hat{\mathbf{w}}}_2^2 > (\tau+2){\hat{w}_\ell}^2-2w_\ell \hat{w}_\ell + {w_\ell}^2 $$
\par Taking the minimum on the right side with respect to $\hat{w}_\ell$ will lead once again to finding the minimum value of a quadratic function. The minimum is found for $\hat{w}_\ell= \frac{w_\ell}{\tau+2}$ and equals to ${w_\ell}^2(1-\frac{1}{\tau+2})$; hence
$$\norm{\mathbf{w}-\hat{\mathbf{w}}}_2^2 >{w_\ell}^2(1-\frac{1}{\tau+2}) \geq q^2(1-\frac{1}{\tau+2}) $$
which once again contradicts the hypothesis so the proof is completed.
\end{IEEEproof}

\section{Proof of Theorem \ref{th:sza-lms}} \label{ap:sza-lms}
\begin{IEEEproof}
Let us define $\bar{\mathbf{w}}(n)$ as the difference between the estimation $\mathbf{w}(n)$ and the true vector $\mathbf{w}$. Subtracting $\mathbf{w}$ from both sides of the equation (\ref{eq:relaxed}) gives
\begin{equation} \label{eq:update}
\begin{aligned}
\bar{\mathbf{w}}(n+1)&=\mathbf{u}(n)-\mathbf{w} -\rho P_s(\mathbf{w}(n))\\
&=\bar{\mathbf{w}}(n)+\mu e(n) \mathbf{x}(n) -\rho P_s(\mathbf{w}(n))\\
\end{aligned} 
\end{equation}
\par After some calculations, which are the same as in the case of the classical LMS, we have that
\begin{equation}
e(n)\mathbf{x}(n)=-\mathbf{x}^T(n)\mathbf{x}(n)\bar{\mathbf{w}}(n)+v(n)\mathbf{x}(n)
\end{equation}
Taking the mean under the independence assumptions made and given that the observation noise mean is zero will yield
\begin{equation}
\mathbb{E}[e(n)\mathbf{x}(n)]=-\mathbf{R}_x\mathbb{E}[\bar{\mathbf{w}}(n)]
\end{equation}
where $\mathbf{R}_x$ is the autocorrelation matrix of $\mathbf{x}(n)$. Then from equation (\ref{eq:update}) we obtain 
\begin{equation}
\mathbb{E}[\bar{\mathbf{w}}(n+1)]=(\mathbf{I}_N-\mu \mathbf{R}_x)\mathbb{E}[\bar{\mathbf{w}}(n)]-\rho \mathbb{E}[P_s(\mathbf{w}(n))]
\end{equation}
where $\mathbf{I}_N$ is the $N\times N$ identity matrix. Given the bound in (\ref{eq:bounds}) the largest eigenvalue of $\mathbf{I}_N-\mu\mathbf{R}_x$ is less than one. Further the term induced by the penalty is bounded by the vectors $-\rho\mathbf{1}$ and $\rho\mathbf{1}$ where $\mathbf{1}$ is the vector of $\mathbb{R}^N$ whose every element is one. Thus we can conclude that the  $\mathbb{E}[\bar{\mathbf{w}}(n)]$ converges and as a result so does  $\mathbb{E}[\mathbf{w}(n)]$. Therefore the algorithm provided by equation (\ref{eq:relaxed}) converges. The limiting vector cannot be found in a closed form but is guaranteed to be the solution of equation (\ref{eq:bias}).
\end{IEEEproof}

%








\end{document}